\documentclass[journal,10pt]{IEEEtran}

\usepackage{amssymb}
\usepackage{graphicx,psfrag} 
\usepackage{epsfig} 
\usepackage{esdiff} 
\usepackage{amsmath,amssymb} 
\usepackage{wasysym}
\usepackage{xfrac}   
\usepackage{subfigure}  
\usepackage{pst-sigsys} 
\usepackage{algorithm, algorithmic}
\usepackage{bbm}
\usepackage{cite}
\usepackage{mathtools}
\usepackage{url}
\usepackage{marginnote}
\usepackage{framed,lipsum}
\usepackage{mdframed}

\usepackage{setspace}
\usepackage{color}
\usepackage{etoolbox}
\makeatletter
\patchcmd{\@makecaption}
  {\scshape}
  {}
  {}
  {}
\makeatother

\ifCLASSINFOpdf
\else
\fi
\begin{document}
\title{Phase Retrieval From Binary Measurements}
\author{Subhadip~Mukherjee
        and~Chandra~Sekhar~Seelamantula, {\it Senior Member, IEEE}
\thanks{\scriptsize The authors are with the Department
of Electrical Engineering, Indian Institute of Science, Bangalore--560012, India. Phone: +918022932695; Fax: +918023600444; Email: subhadipm@iisc.ac.in, chandra.sekhar@ieee.org. 
}
}

\markboth{}
{Shell \MakeLowercase{\textit{et al.}}: Bare Demo of IEEEtran.cls for Journals}
\maketitle

\begin{abstract}
We consider the problem of signal reconstruction from quadratic measurements that are encoded as $+1$ or $-1$ depending on whether they exceed a predetermined positive threshold or not. Binary measurements are fast to acquire and inexpensive in terms of hardware. We formulate the problem of signal reconstruction using a \textit{consistency criterion}, wherein one seeks to find a signal that is in agreement with the measurements. To enforce consistency, we construct a convex cost using a one-sided quadratic penalty and minimize it using an iterative accelerated projected gradient-descent (APGD) technique. The PGD scheme reduces the cost function in each iteration, whereas incorporating momentum into PGD, notwithstanding the lack of such a descent property, exhibits faster convergence than PGD empirically. We refer to the resulting algorithm as \textit{binary phase retrieval} (BPR). Considering additive white noise contamination prior to quantization, we also derive the Cram\'er-Rao Bound (CRB) for the binary encoding model. Experimental results demonstrate that the BPR algorithm yields a signal-to-reconstruction error ratio (SRER) of approximately $25$ dB in the absence of noise. In the presence of noise prior to quantization, the SRER is within $2$ to $3$ dB of the CRB.             
\end{abstract}
\begin{IEEEkeywords}
Binary phase retrieval, consistency, lifting, accelerated projected gradient-descent, Cram\'er-Rao bound.
\end{IEEEkeywords}

\section{Introduction}
\IEEEPARstart{P}{hase} retrieval (PR) is encountered in several imaging applications such as X-ray crystallography \cite{crystallography}, holography \cite{holography}, microscopy \cite{microscopy}, coherent modulation imaging \cite{zhang_chen_nature}, coherent diffraction imaging \cite{shen_tan, shen_bao}, etc. Since the sensors can record only complex wavefront intensities, it becomes imperative to recover the phase from the magnitude measurement in order to reconstruct the underlying object. This ill-posed inverse problem can be solved by acquiring oversampled magnitude measurements and incorporating signal priors such as non-negativity, compact support, sparsity, etc.\\
\indent The early contributions in PR were due to Fienup \cite{fienup_main1,bauschke}, Gerchberg and Saxton \cite{gs}, who proposed iterative error reduction algorithms. There are also techniques that rely on Hilbert integral relations between the log-magnitude and the phase of the Fourier transform for certain classes of signals \cite{css_holography}. Recently, we developed PR algorithms for a class of two-dimensional (2-D) signals \cite{param2D_tip} and for signals belonging to principal shift-invariant spaces \cite{pr_shift_invar}.\\ 
\indent Recently, the problem of PR has been addressed within the realm of sparsity and magnitude-only \textit{compressive sensing} (CS). Yu and Vetterli have proposed a sparse spectral factorization technique \cite{yu_vetterli}, and established uniqueness guarantees. Moravec et al. proposed \textit{compressive PR} \cite{moravec_cpr}, by enforcing the constraint of compressibility on the signal. A greedy local search-based algorithm for sparse PR (GESPAR), was proposed by Schechtman et al. \cite{gespar}. Other notable contributions for sparse PR include techniques based on dictionary learning (DOLPHIn) \cite{dolphin}, generalized message passing \cite{schniter_gamp}, simulated annealing for sparse Boolean signals  \cite{bspr_peng}, majorization-minimization for recovery from undersampled measurements \cite{palomar_mm}, etc. We developed the {\it sparse Fienup algorithm} \cite{css_fienup}, where sparsity is enforced via hard-thresholding in the signal domain. Vaswani et al. \cite{low_rank_PR_vaswani} recently proposed an alternating minimization technique for  recovering a low-rank matrix from quadratic measurements corresponding to projections with each of its columns. Fogel et al. \cite{fogel_pr_imaging} showed that incorporating signal priors such as sparsity and positivity lead to a significant speed-up of iterative reconstruction techniques.\\    
\indent A seminal contribution in PR is the \textit{PhaseLift} framework of Cand\`es et al. \cite{PL1,PL2}, which relies on lifting the ground-truth vector to a matrix such that the quadratic measurements get converted to an equivalent set of linear measurements. Reconstruction is achieved by solving a tractable semi-definite program (SDP). Sparsity was imposed within the PhaseLift framework using the $\ell_1$ penalty \cite{cpr_pl}, or \textit{log-det} relaxation \cite{qcs_eldar}. Gradient-descent approaches for PR that do not rely on lifting include the Wirtinger Flow (WF) method \cite{candes_wf_it} and its truncated version (TWF) \cite{candes_twf}. These algorithms are scalable and have convergence guarantees for the spectral initialization \cite{netrapalli}. Waldspurger et al. developed \textit{PhaseCut} \cite{phasecut}, where PR is formulated as a non-convex quadratic program and solved using a block-coordinate-descent approach, having a per-iteration complexity comparable to that of Gerchberg-Saxton-type algorithms.\\
\indent The problem of measurement quantization was considered in the context of CS, but not PR. Zymnis et al. considered the problem of reconstruction from quantized CS measurements \cite{zymnis_qcs}. Boufounos and Baraniuk addressed the problem of binary CS \cite{boufounous} and proposed a fixed-point continuation algorithm for signal recovery. The other notable works in the context of binary CS include \cite{nowak1,plan1,msp,aop,biht,plan2,binary_imaging2}.\\
\textit{This Paper}: We consider a scenario where quadratic measurements of a signal are compared with a threshold $\tau>0$ and are encoded using the binary alphabet $\pm 1$. From the perspective of analog-to-digital conversion, it is efficient to encode coarsely by sampling at a high rate, than to encode finely at a low sampling rate \cite{texasinstru,motorola,walden,binle}. Our reconstruction algorithm combines the principles of lifting and consistent reconstruction, originally introduced in \cite{boufounous}, and employs an accelerated projected gradient-descent strategy, which is found to have a better empirical performance than PGD (cf. Appendix~\ref{pgd_proof}). We also consider additive white noise contamination before binary encoding and derive the Cram\'er-Rao Bound (CRB), which serves as the theoretical benchmark. Experimental results demonstrate that the proposed algorithm yields a reconstruction that is accurate to within $2$ to $3$ dB of the CRB. 
\section{The Binary Phase Retrieval (BPR) problem}
\indent The objective in standard PR for real signals is to reconstruct $\boldsymbol x^* \in \mathbb{R}^n$ from quadratic measurements $b_i=\left| \boldsymbol a_i^\top \boldsymbol x^*  \right|^2, i=1:m$, where $\{\boldsymbol a_i\}$ are Gaussian sampling vectors drawn independently from $\mathcal{N}\left(\boldsymbol 0,\boldsymbol I_n\right)$, $\boldsymbol I_{n}$ being the $n \times n$ identity matrix. The notation $i=1:m$ is used as a compact version of $i = 1, 2, \cdots, m$. In BPR, the squared-magnitude measurements are encoded using $-1$ or $+1$ by comparing them against a predetermined threshold $\tau>0$, resulting in the sign measurements $y_i=\text{sgn}\left(\left| \boldsymbol a_i^\top \boldsymbol x^*  \right|^2 -\tau\right), i = 1 : m$, where $\text{sgn}(\cdot)$ denotes the signum function.\\ 
\indent The key idea behind \textit{consistent reconstruction} is to seek a vector $\boldsymbol x$ that is in agreement with the measurements, so that the reconstructed vector, when passed through the same acquisition process, matches the given measurements. The consistency condition could be expressed succinctly as $y_i\left(\left| \boldsymbol a_i^\top \boldsymbol x  \right|^2-\tau\right)>0, \forall i$. Effectively, the  problem is:
\begin{equation}
\text{Find\,\,} \boldsymbol x \in \mathbb{R}^n \text{\,\,s.t.\,\,} y_i\left(\left| \boldsymbol a_i^\top \boldsymbol x  \right|^2-\tau\right)>0, i=1:m.
\label{bpr_consistency_req}
\end{equation}
\indent We combine the requirement of \textit{consistent recovery} with the principle of \textit{lifting} \cite{PL1,PL2}, and formulate a suitable cost function for minimization. By \textit{lifting}, one expresses the quadratic term as $\left| \boldsymbol a_i^\top \boldsymbol x  \right|^2=\text{Tr}\left(\boldsymbol A_i \boldsymbol X\right)$, where $\boldsymbol X = \boldsymbol x \boldsymbol x^\top$, $\boldsymbol A_i = \boldsymbol a_i \boldsymbol a_i^\top$, and $\text{Tr}(\cdot)$ denotes the trace operator. Since $\boldsymbol X$ is positive semi-definite ($\boldsymbol X\succeq \boldsymbol 0$) and has rank one, consistent recovery in the lifted domain takes the following form:
\begin{equation*}
\text{Find\,\,} \boldsymbol X\succeq \boldsymbol 0 \text{\,\,s.t.\,\,} y_i\left(\text{Tr}\left(\boldsymbol A_i \boldsymbol X\right)-\tau\right)>0 \text{\,\,and\,\,}\text{rank}(\boldsymbol X)=1,
\label{sdpr_lifted}
\end{equation*}
for $i=1:m$.~In order to solve this problem, we formulate an optimization cost using the one-sided quadratic loss $f:\mathbb{R}\rightarrow \mathbb{R}$, defined as 
$f(u) = \frac{1}{2}u^2\mathbbm{1}_{(u\leq0)}$, where $\mathbbm{1}$ is the indicator function. 
The BPR problem is cast as
\begin{equation}
\hat{\boldsymbol X}=\arg\underset{\boldsymbol X \succeq 0}{\min}\text{\,} F(\boldsymbol{X}) \text{\,\,subject to\,\,}\text{rank}\left({\boldsymbol X}\right)=1, 
\label{uc_recovery_sdp}
\end{equation}
where $F\left(\boldsymbol X \right)=\sum_{i=1}^{m}f\left(y_i\left(\text{Tr}\left(\boldsymbol A_i \boldsymbol X\right) -\tau\right)\right)$. The one-sided quadratic loss essentially penalizes {\it lack of consistency}. The problem in \eqref{uc_recovery_sdp} can be solved by employing projected gradient-descent (PGD) or its accelerated counterpart (APGD), which incorporates a momentum factor \cite{nesterov1}. Our recovery algorithm for BPR employing APGD is listed in Algorithm~\ref{algo_sdpr}. The rank-1 projection can be computed efficiently using {\it power iterations} \cite[Ch. 7]{power_it}. A proof that the PGD scheme decreases the cost in each iteration is given in Appendix~\ref{pgd_proof}. This guarantee does not carry over to APGD because although the cost in \eqref{uc_recovery_sdp} is convex, the rank-1 constraint is not. However, experimentally, we found that APGD leads to a faster convergence and hence we employ APGD in the proposed BPR algorithm. Similar observations were made in the context of low-rank matrix completion \cite{geng_yang} and PhaseLift \cite{PL2}.
\begin{algorithm}[t]
\caption{The Binary Phase Retrieval (BPR) algorithm.}
\begin{algorithmic}
\STATE {\bf  1.} {\bf Initialization}: Set $\boldsymbol X^{0}=\boldsymbol Y^{0} = \boldsymbol 0_{n\times n}$, $\theta^0 = 1$, and $N_{\text{iter}}$ = Maximum iteration count.
\STATE {\bf  2.} {\bf For} $t=1:N_{\text{iter}}$, {\bf do}:
\begin{enumerate}
\item Line-search: $\eta^t=\arg\underset{\eta>0}{\min}\text{\,\,}F\left(\boldsymbol X^t-\eta\boldsymbol \nabla F\left(\boldsymbol X^t\right)\right)$,
\item ${\boldsymbol X}^{t+1}=\mathcal{P}_{\text{rank}-1}\left({\boldsymbol Y}^{t}-\eta^t  \nabla F\left( \boldsymbol Y^t\right)\right) = \hat{\boldsymbol x}\hat{\boldsymbol x}^\top$,
\item $\theta^{t+1}=2\left(1+\sqrt{1+\frac{4}{\left(\theta^{t}\right)^2}} \right)^{-1}$, and
\item $\boldsymbol Y^{t+1}={\boldsymbol X}^{t+1}+\theta^{t+1} \left(\frac{1}{\theta^t}-1 \right)\left({\boldsymbol X}^{t+1}-{\boldsymbol X}^{t}\right)$. \end{enumerate}
\STATE {\bf  3.} {\bf Output}: $\hat{\boldsymbol x}$, which is an estimate of $\boldsymbol x^*$.
\end{algorithmic}
\label{algo_sdpr}
\end{algorithm}

\section{Simulation Results}
\indent If $\hat{\boldsymbol x}\in \mathbb{R}^n$ is a consistent solution to the BPR problem, so is $-\hat{\boldsymbol x}$. In order to factor out the effect of the global sign, an appropriate measure to quantify the accuracy of reconstruction {\it vis-\`a-vis} the ground truth $\boldsymbol x^*$ would be the globally-sign-invariant signal-to-reconstruction error ratio (SRER) \cite{PL2}, defined as $\text{SRER}=10\log_{10}\left[\underset{\alpha\in\{-1,+1\}}{\max}\frac{\left\| \boldsymbol x^* \right\|_2^2}{\left\| \alpha\,\hat{\boldsymbol x}-\boldsymbol x^* \right\|_2^2}\right] \text{dB}$. The second measure that is relevant in the context of BPR is \textit{consistency} of the reconstruction $\hat{\boldsymbol x}$ with the measurements, which is computed as $\Upsilon=\frac{1}{m}\displaystyle\sum_{i=1}^{m}\mathbbm{1}_{\left(y_i\left(\left| \boldsymbol a_i^\top \hat{\boldsymbol x}  \right|^2-\tau\right)>0 \right)}$. The consistency measure $\Upsilon$ is the fraction of measurements correctly explained by the reconstruction $\hat{\boldsymbol x}$. Consequently, $0 \leq \Upsilon \leq 1$, and $\Upsilon = 1$ is the best one could hope to achieve.
\begin{figure}[t]
\centering
$
\begin{array}{cc}
\includegraphics[width=1.65in]{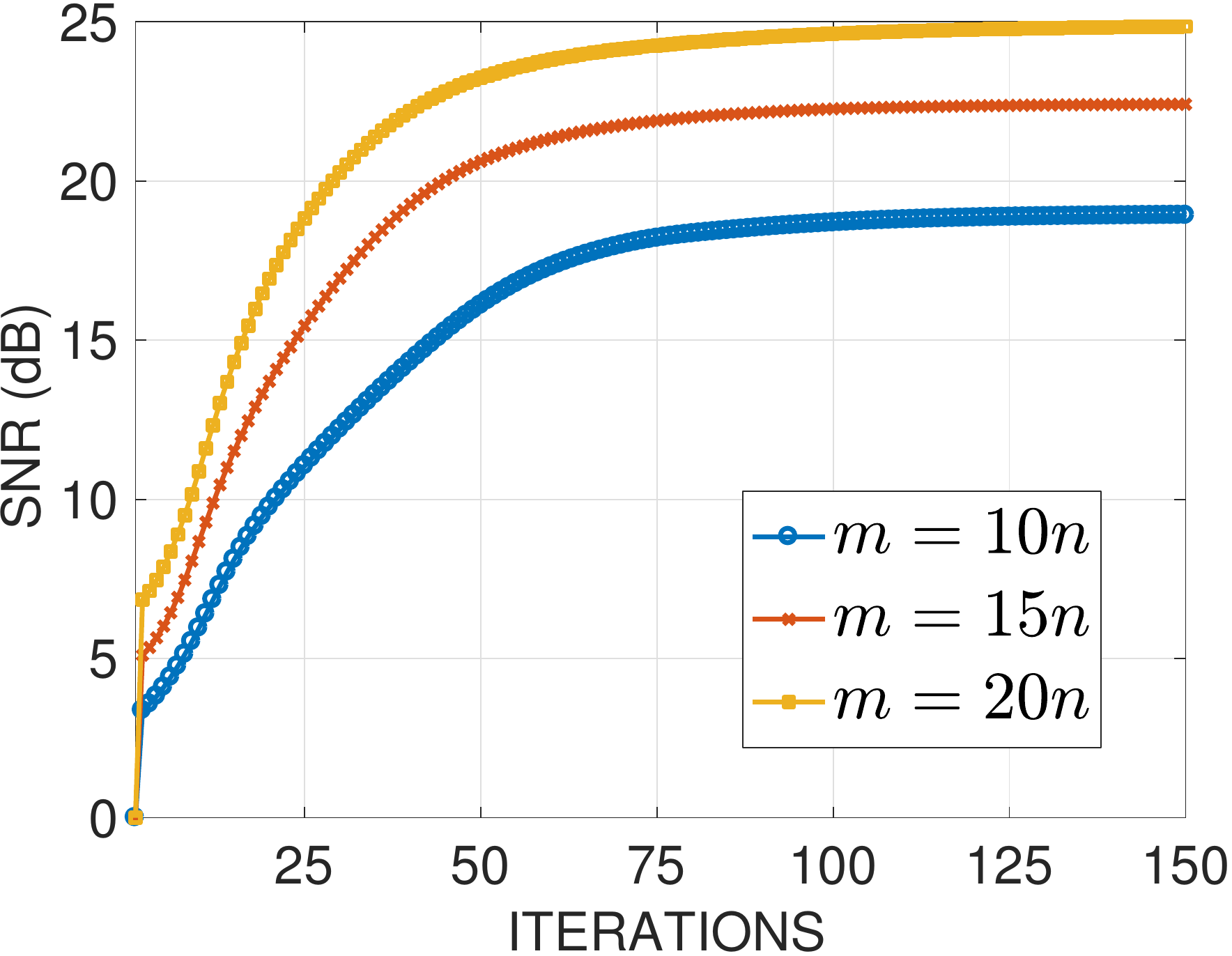}&
\includegraphics[width=1.65in]{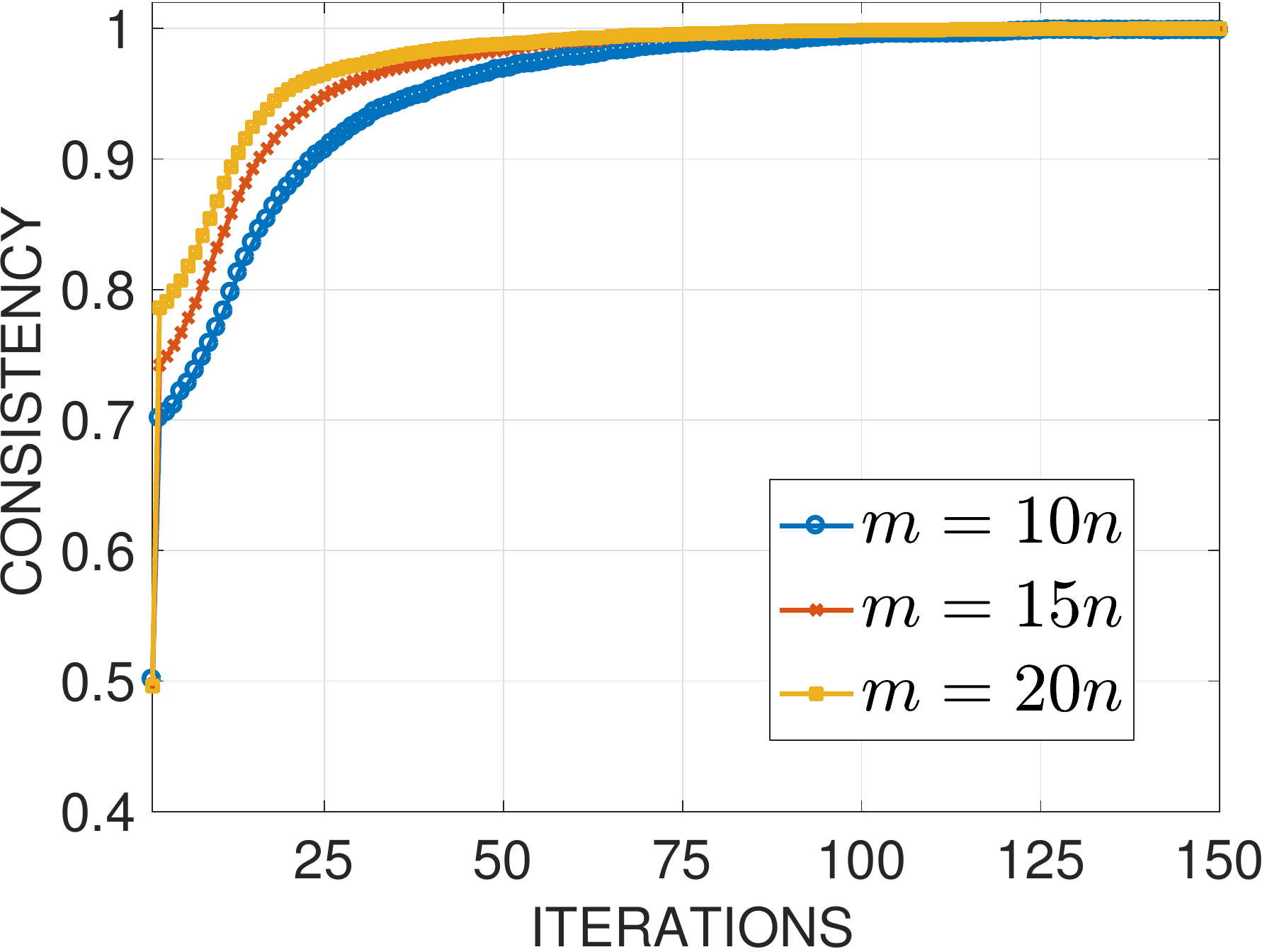}\\
\small {\text{(a)}}&\small {\text{(b)}}\\
\includegraphics[width=1.6in]{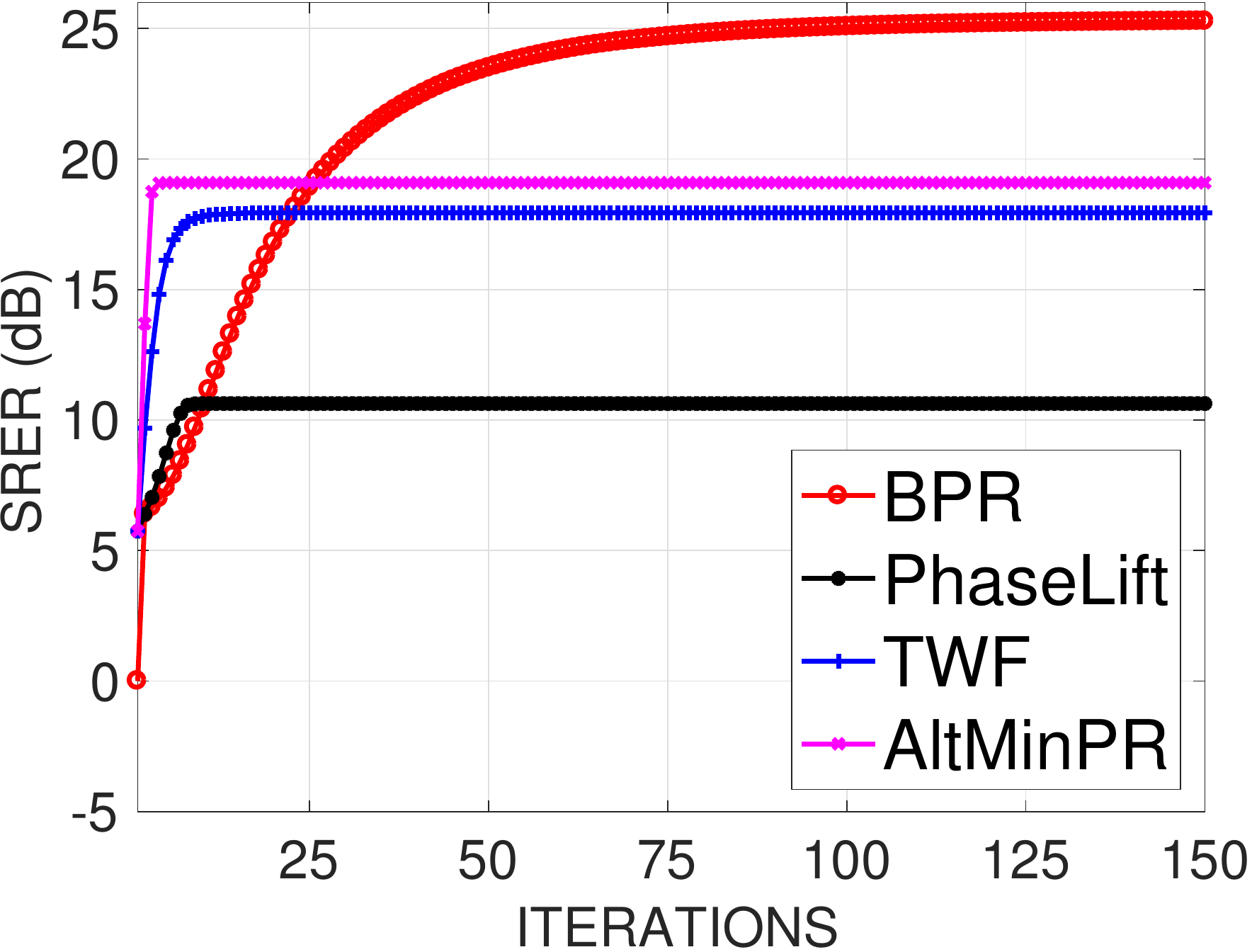}&
\includegraphics[width=1.6in]{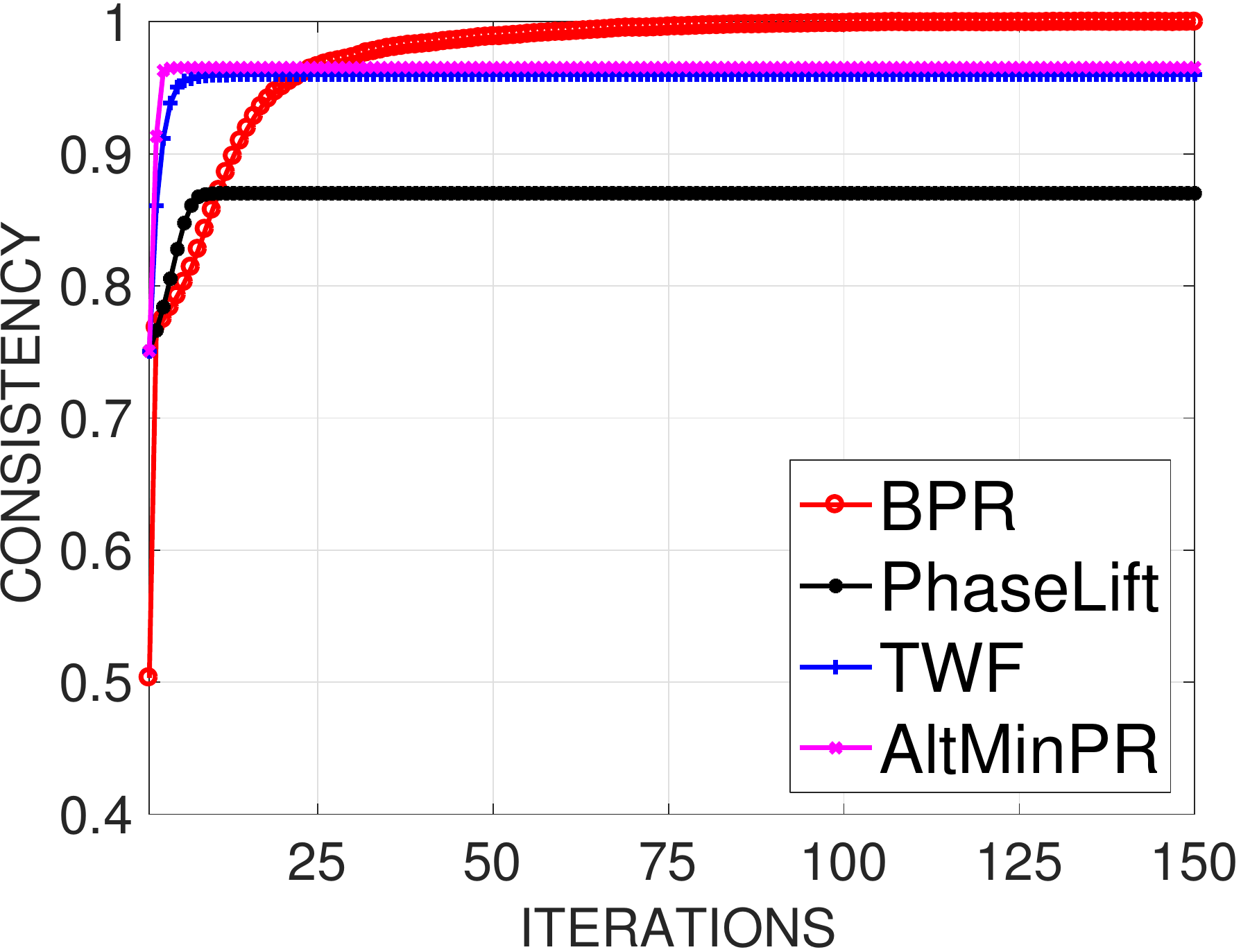}\\
\small {\text{(c)}}&\small {\text{(d)}}
\end{array}
$
\caption{\small{(Color online) Performance assessment of the BPR algorithm on noise-free measurements: (a) Reconstruction SRER; and (b) Consistency, for different values of $m$. A comparison of (c) SRER; and (d) consistency \textit{vis-\`a-vis} the state-of-the-art algorithms for $m=20n$.}}
\label{BPR_noiseless_fig}
\end{figure}
\begin{figure}[t]
\centering
$
\begin{array}{ccc}
\includegraphics[width=1.5in]{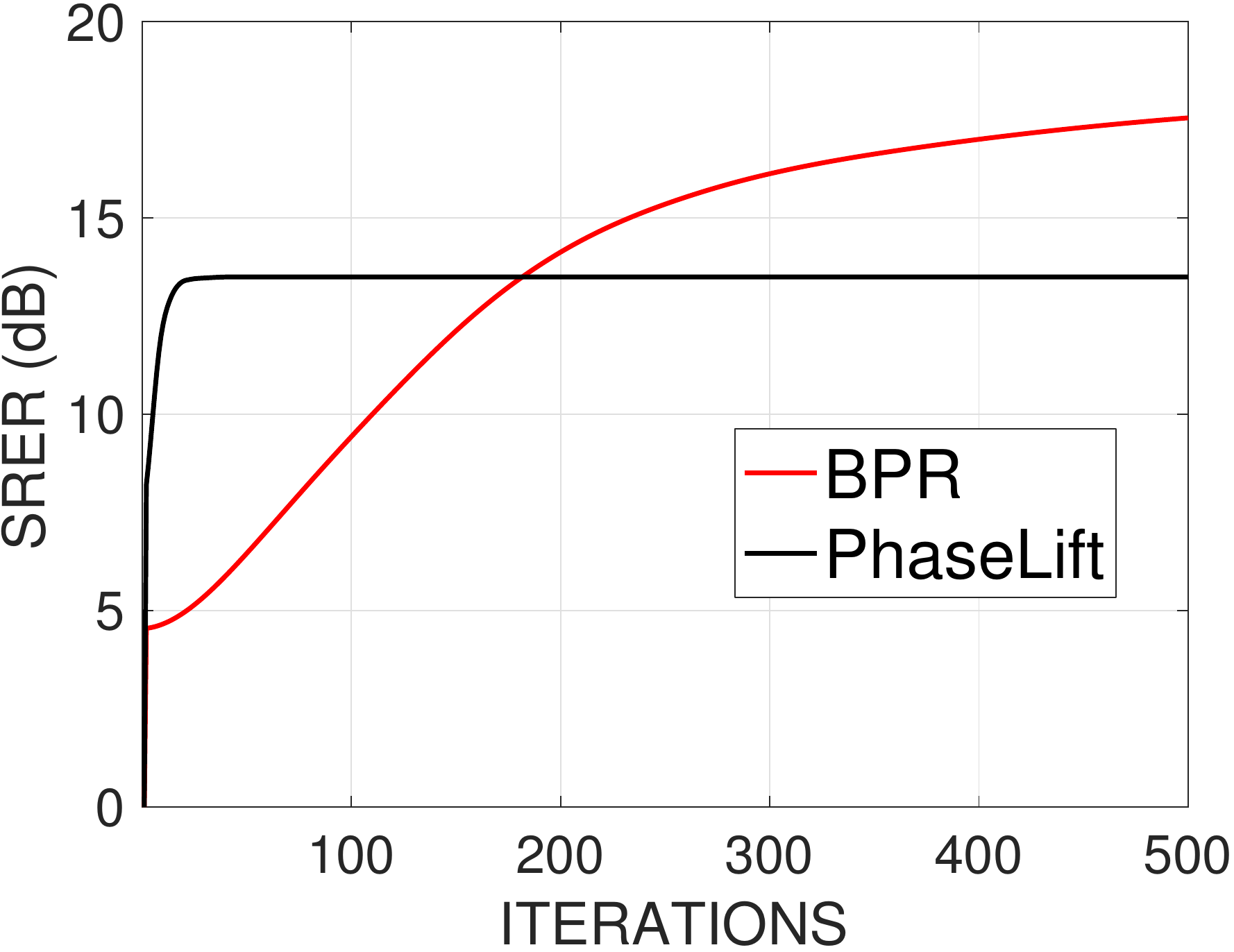}&
\includegraphics[width=1.5in]{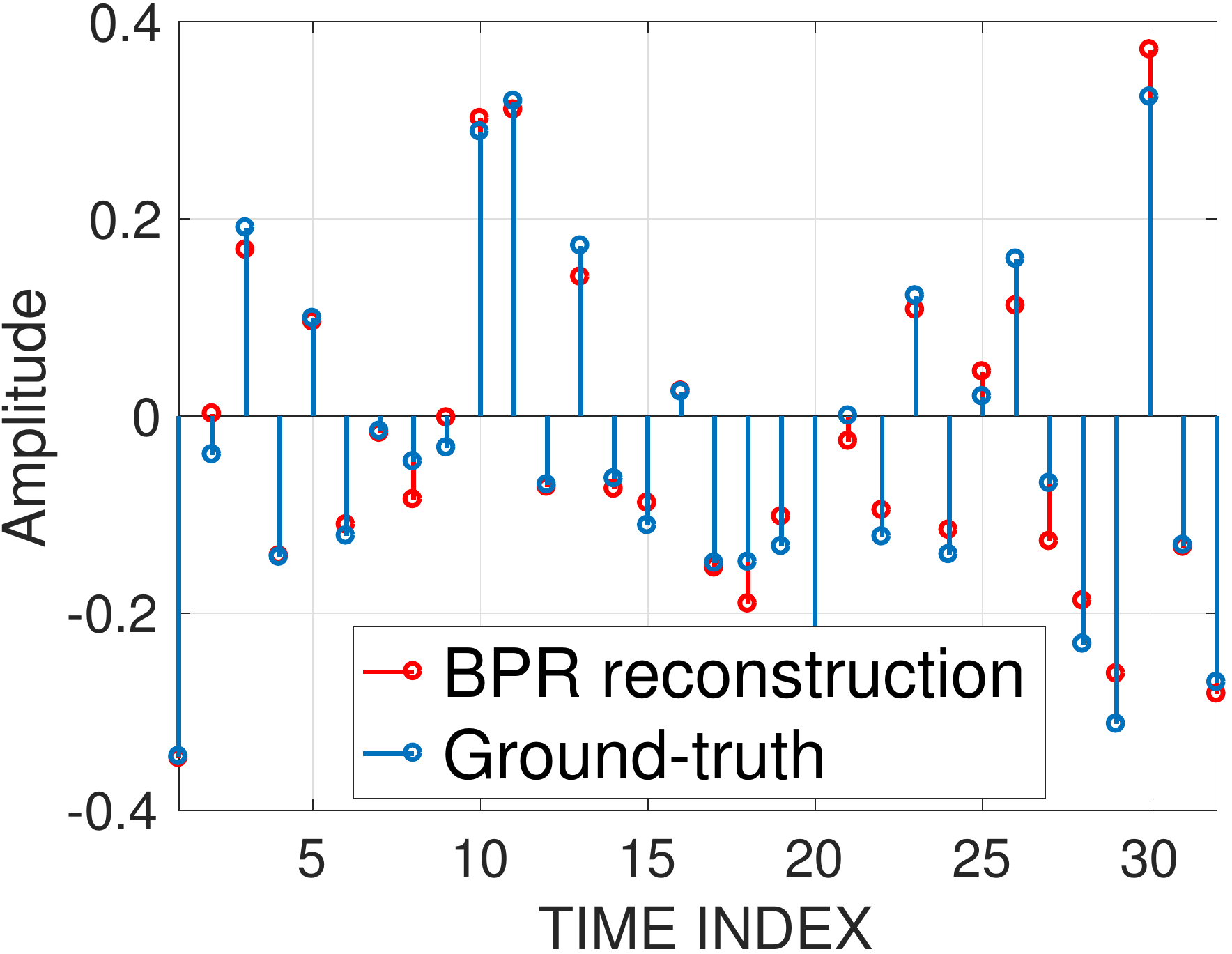}\\
\text{\small (a) SRER versus iterations} & \text{\small (b) BPR reconstruction} \\
\end{array}
$
\caption{\small{(Color online) A comparison of BPR and PhaseLift corresponding to structured illumination with an oversampling of $\frac{m}{n}=20$.}}
\label{BPR_fourier_fig1}
\end{figure}
\reversemarginpar
\subsection{Signal Reconstruction in the Absence of Noise}
\label{noiseless_synthetic_exp_sec}
\indent Consider an instance of $\boldsymbol x^*$ drawn uniformly at random from the unit-sphere in $\mathbb{R}^n$, and the measurement vectors $\boldsymbol a_i \sim \mathcal{N}\left(\boldsymbol 0,\boldsymbol I_{n}\right)$, where $n=64$. The threshold $\tau$ is chosen such that the measurements are encoded as $+1$ or $-1$ with equal probability. In this case, $|\boldsymbol{a}_i^\top \boldsymbol{x}^*|^2$ follows a $\chi^2_1$ distribution, corresponding to which the threshold value turns out to be $\tau=0.4550$. The equiprobable encoding strategy is popular and was also adopted in quantized CS \cite{zymnis_qcs} and binary CS problems \cite{boufounous}. The optimal step-size $\eta^t$ is determined according to Step 2.1 of Algorithm~\ref{algo_sdpr}, with a small search range $[0,0.0025]$ and precision of $10^{-5}$.\\
\indent The SRER and consistency versus iterations corresponding to different oversampling factors $\frac{m}{n}$ are shown in Figures~\ref{BPR_noiseless_fig}(a) and (b), respectively. The results have been averaged over $20$ independent trials. As expected, the SRER increases with increase in $\frac{m}{n}$. Higher oversampling factors also lead to faster convergence of $\Upsilon$.\\
\indent To the best of our knowledge, this paper introduces the binary PR problem for the first time. Hence, there is no prior art for making comparisons. One way to compare with techniques such as PhaseLift \cite{PL1,PL2}, AltMinPR \cite{netrapalli}, and TWF \cite{candes_twf}, is to model the quantization noise as an additive perturbation on the measurements. Further, such a comparison calls for an appropriate encoding of $y_i$ for the competing techniques -- the $\pm 1$ encoding would not be appropriate for them because their cost functions involve a quadratic that measures the distance between $|\boldsymbol{a}_i^\top \hat{\boldsymbol{x}}|^2$ and $y_i$, and no $ \hat{\boldsymbol{x}}$, not even ${\boldsymbol{x}^*}$, would optimize their cost function. This is not an issue with BPR since the cost relies on consistency. Hence, in order to be fair to the other techniques, we replace the $-1$ and $+1$ symbols with the centroids of the intervals $[0,\tau]$ and $[\tau, \infty)$, respectively, computed with respect to the $\chi_1^2$ density. These turn out to be $0.1427$ and $1.8573$, respectively. The details of the settings for the competing algorithms are provided in Appendix~\ref{other_algo_settings}.\\
\indent A comparison is shown in Figures~\ref{BPR_noiseless_fig}(c) and (d) for the same experimental setup considered in Figures~\ref{BPR_noiseless_fig}(a) and (b). The competing techniques converge relatively fast and do a reasonable job even with binary quantization. The BPR algorithm, on the other hand, takes more iterations to ensure high consistency, but ultimately results in an estimate that has a much higher accuracy (about 5 dB in this instance) and superior consistency with the measurements. A comparison of the run-times is provided in Appendix~\ref{run_time_sec}.
\subsection{Signal Reconstruction With Fourier Measurements}
\indent Although Gaussian measurements are considered in Section~\ref{noiseless_synthetic_exp_sec}, the BPR algorithm can be applied to Fourier measurements as well. Consider a Fourier sampling scheme of the structured illumination type, employed in the context of PhaseLift \cite{PL2}. In this setup, one considers the measurement matrix $\boldsymbol A = \left[  (\boldsymbol F  \boldsymbol W_1)^\top (\boldsymbol F  \boldsymbol W_2)^\top \cdots (\boldsymbol F  \boldsymbol W_k)^\top\right]^\top$
where $\boldsymbol F$ is the $n \times n$ discrete Fourier transform (DFT) matrix, $\boldsymbol W_j$s are $n \times n$  diagonal matrices containing random binary entries (0 or 1 with probability $\frac{1}{2}$) on the diagonal, and $k=\frac{m}{n}$ is the oversampling factor. The measurements $|\boldsymbol{a}_i^\top\boldsymbol x^*|^2$, where $\boldsymbol{a}_i$ denotes the $i^{\text{th}}$ row of $\boldsymbol{A}$, are quantized as $\pm 1$, depending on whether they exceed a threshold $\tau$ or not. The threshold $\tau$ is set such that $\text{Prob}\left(|\boldsymbol{a}_i^\top\boldsymbol x^*|^2 > \tau \right) = \text{Prob}\left(|\boldsymbol{a}_i^\top\boldsymbol x^*|^2 < \tau \right) = \frac{1}{2}$. The reconstruction performance of BPR and PhaseLift for this setting is shown in Figure~\ref{BPR_fourier_fig1}. We observe that PhaseLift converges faster than BPR, but the SRER of BPR, upon convergence, is about $4$ dB higher than that of PhaseLift. Further details of BPR with Fourier measurements are provided in Appendix~\ref{bpr_dft_app}.
\subsection{Signal Reconstruction in the Presence of Noise}
\label{bpr_measurement_noise_exp_sec}
\begin{figure}[t]
$
\begin{array}{ccc}
\centering
		\subfigure[]{\includegraphics[width=1.075in]{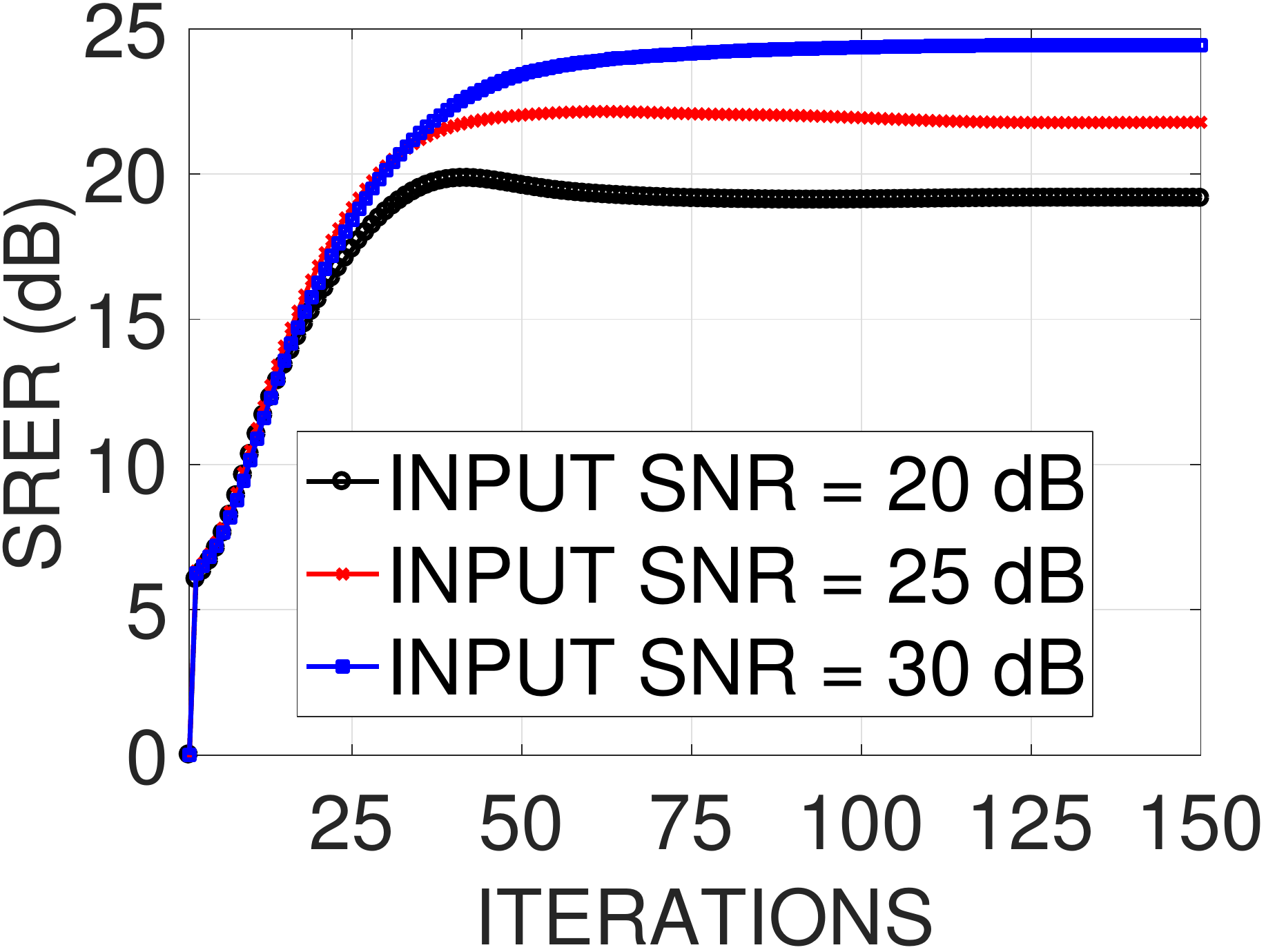}}&
		\subfigure[]{\includegraphics[width=1.075in]{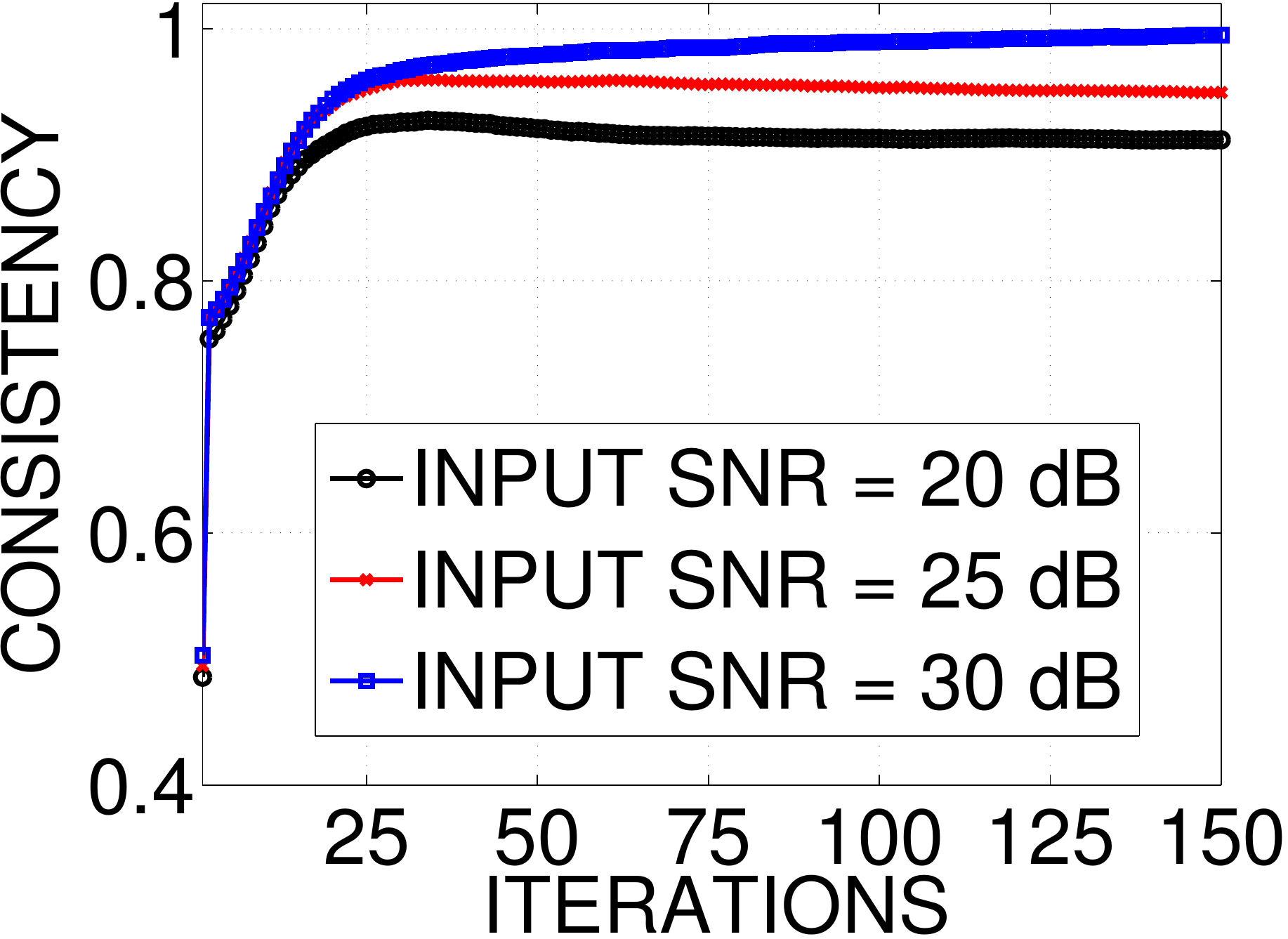}}&
		\subfigure[]{\includegraphics[width=1in]{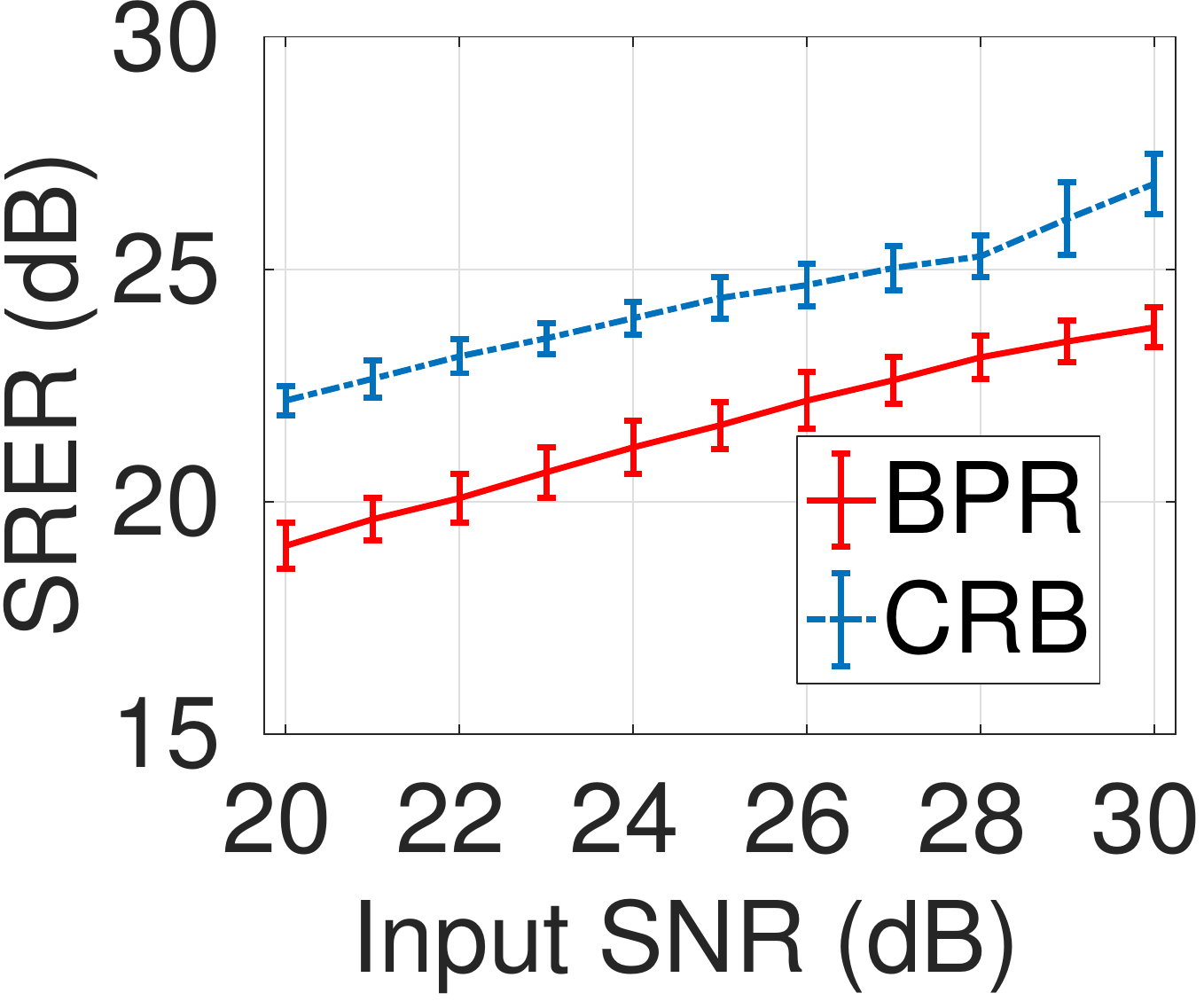}}
\end{array}
$
\caption{\small{(Color online) Performance of BPR in noise: (a) reconstruction SRER, (b) consistency, and (c) SRER versus the CRB.}}
\label{BPR_noisy_fig}
\end{figure}

\indent The measurements in the presence of additive white noise $\xi_i$ prior to quantization are given by
\begin{equation}
y_i=\text{sgn}\left(\left| \boldsymbol a_i^\top \boldsymbol x^*  \right|^2+\xi_i -\tau\right), i = 1 : m,
\label{measurement_eq_binary_noisy}
\end{equation} 
where $\{\xi_i\}_{i=1}^{m}\stackrel{\text{i.i.d.}}{\sim} \mathcal{N}\left(0,\sigma_{\xi}^2\right)$. The input SNR is defined as $\text{SNR}_\text{in}=\frac{1}{m\sigma_{\xi}^2}\sum_{i=1}^{m}\left| \boldsymbol a_i^\top \boldsymbol x^*  \right|^4$. The experimental parameters are kept the same as in Section~\ref{noiseless_synthetic_exp_sec} with $m=20n$. The results are shown in Figure~\ref{BPR_noisy_fig}. From Figure~\ref{BPR_noisy_fig}(a), we observe that the SRER steadily improves with increasing input SNR. A comparison with Figure~\ref{BPR_noiseless_fig}(a) reveals that the SRER corresponding to $\text{SNR}_\text{in}=30\text{\,dB}$ is nearly the same as that obtained with clean measurements. The consistency is also high (cf. Figure~\ref{BPR_noisy_fig}(b)), which is indicative of the inherent noise robustness due to binary quantization. As expected, the consistency drops at low input SNR.

\subsection{Noise Robustness: SRER vis-\`a-vis the CRB}
\label{np_crb_sec}
\indent The theoretical benchmark against which the performance of the BPR algorithm could be compared is the CRB, which is derived in Appendix A. For illustration, we consider the ground-truth signal $\boldsymbol x^*$ to be a sum of two sinusoids, with the $\ell^{\text{th}}$ entry $x_{\ell}^*=\kappa\left[1.5 \sin\left(\frac{4\pi\ell}{n}\right) + 2.5 \cos\left(\frac{14\pi\ell}{n}\right)\right], \ell=0:n-1$, where $n=64$ and the normalizing constant $\kappa$ ensures that $\left\|\boldsymbol x^*\right\|_2=1$. The sampling vectors $\left\{\boldsymbol a_i\right\}_{i=1}^{m} \sim \mathcal{N}\left(\boldsymbol 0,\boldsymbol I_n\right)$. Reconstruction is carried out using BPR and the SRER corresponding to each input SNR is averaged over $20$ independent noise realizations for a fixed set of ${\boldsymbol a}_i$. Since $\left\{\boldsymbol a_i\right\}_{i=1}^{m}$ are random, we have to perform one more level of averaging of the SRERs with respect to the realizations of ${\boldsymbol a}_i$. For this purpose, we generate $20$ different measurement matrices and compute the average SRER and the standard deviation -- these are shown in Figure~\ref{BPR_noisy_fig}(c) as a function of the input SNR. We observe that BPR attains SRERs within $2$ to $3$ dB of the CRB at all input SNRs. In this case, the standard deviations are limited to within $1$ dB of the average, which goes to show that the variability in SRER with respect to $\left\{\boldsymbol a_i\right\}$ is small.  
\subsection{An Example of Image Reconstruction}
\begin{figure}[t]
$
\begin{array}{cc}
	\centering
	\subfigure[Ground-truth]{\includegraphics[width=1.50in]{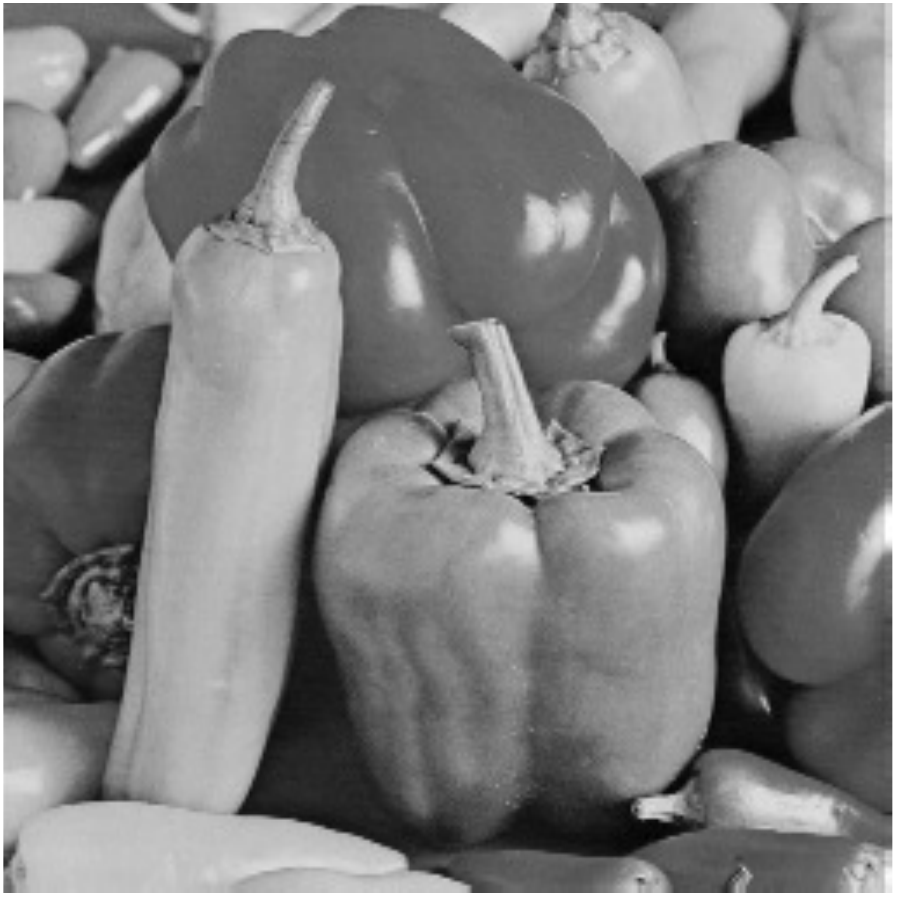}} &
	\subfigure[BPR reconstruction]{\includegraphics[width=1.50in]{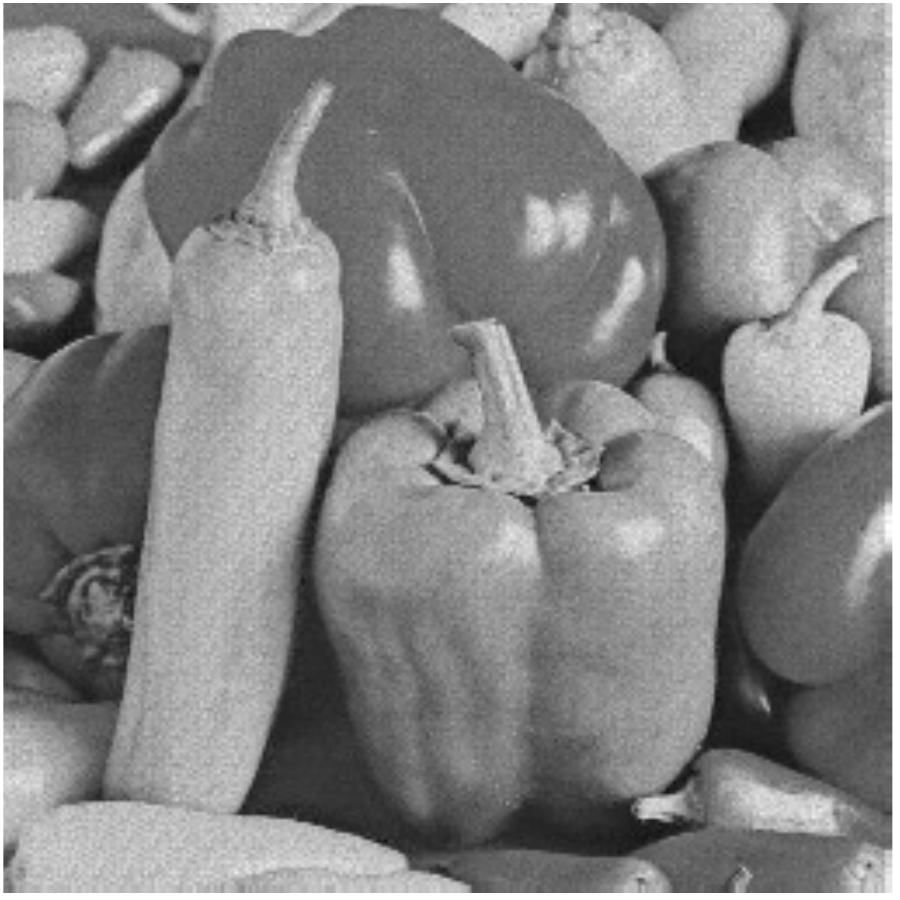}}\\
	\subfigure[Difference image]{\includegraphics[width=1.50in]{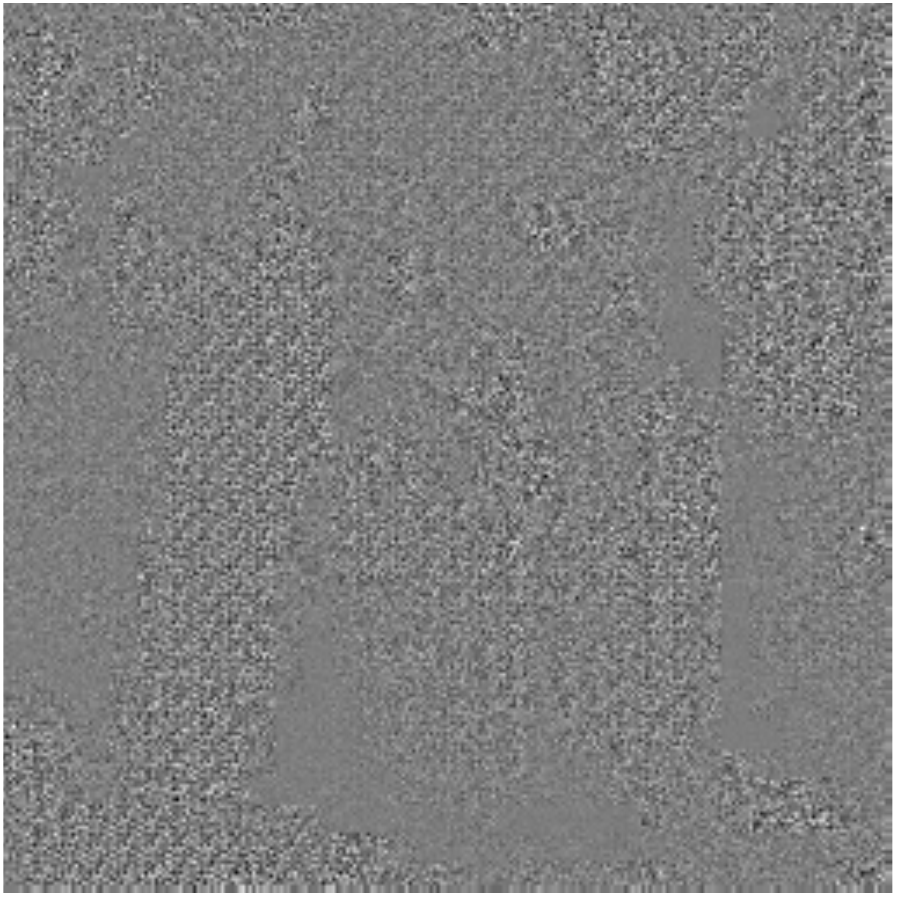}}&
	\subfigure[PSNR and SSIM vs. $\frac{m}{n}$]{\includegraphics[height=1.34in]{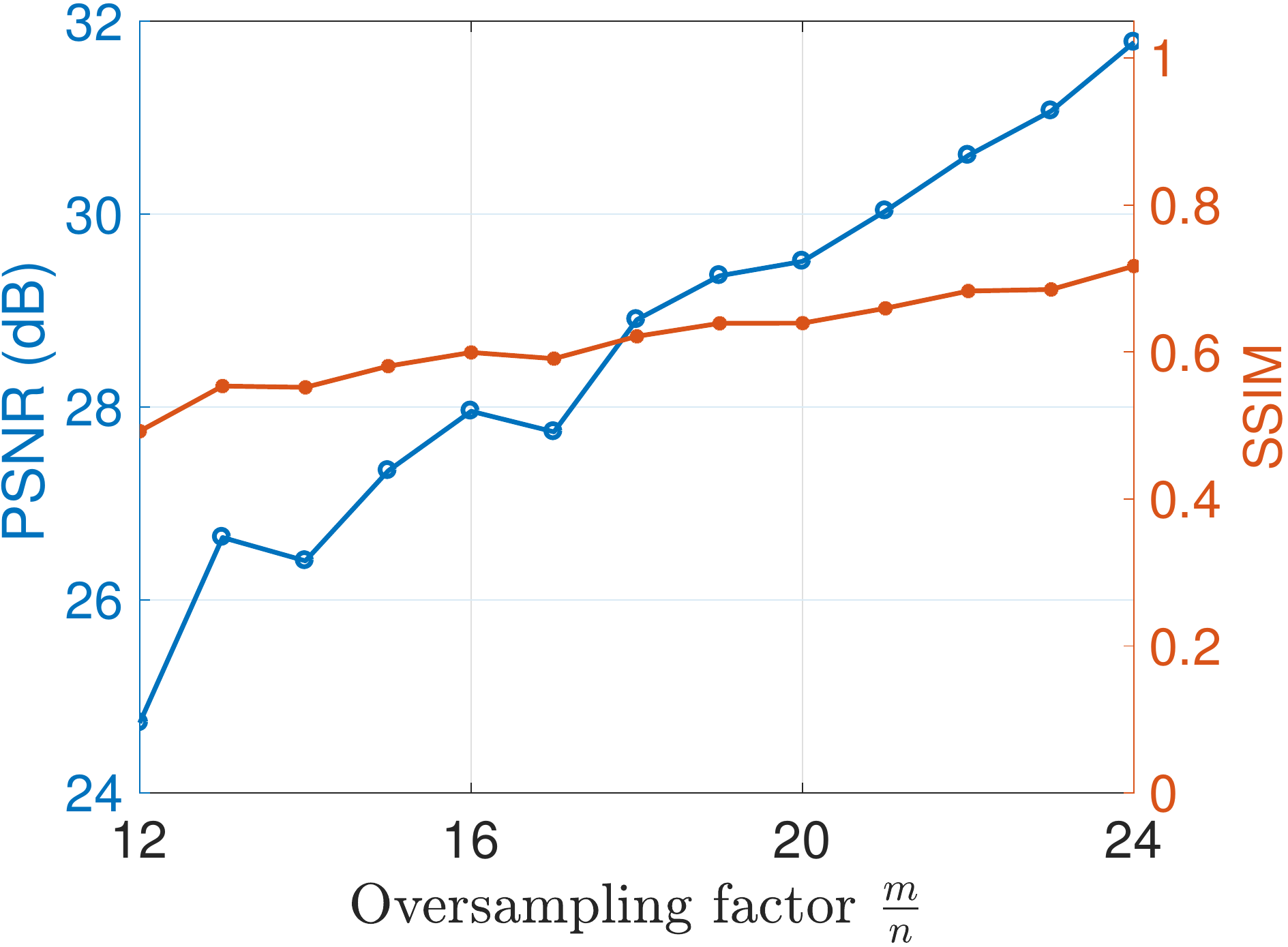}}
\end{array}
$
\caption{\small{(Color online) Image reconstruction using BPR. The reconstruction shown in (b) has PSNR = 29.60 dB and SSIM = $0.64$.}}
\label{sdpr_image_recon}
\end{figure}

\indent Consider the \emph{Peppers} image of size $256\times256$ (cf. Figure~\ref{sdpr_image_recon}(a)), divided into nonoverlapping patches of size $8 \times 8$ leading to a total of $1024$ patches. The effective dimension of the image is $n = 256^2$ and the {\it total} number of measurements is $m$. The sampling vectors are drawn independently following $\boldsymbol a_i\sim \mathcal{N}(0,\boldsymbol I_{64}), 1 \leq i \leq m/1024$. We analyze the reconstruction performance ($N_{\text{iter}}=75$) as a function of the oversampling factor $\frac{m}{n}$. Reconstruction is performed patch-wise. A small search range $[0,0.0055]$ is chosen for the optimal $\eta^t$ with a precision of $10^{-5}$. The image reconstruction quality is quantified using the structural similarity index (SSIM) \cite{ssim_ref} and the peak SNR defined as $\text{PSNR}=20\log_{10}\frac{255\sqrt{n}}{\left\|\boldsymbol I - \hat{\boldsymbol I}\right\|_{\textsc{F}}} \text{\,\,dB}$, where $\boldsymbol{I}$ is the image, $\hat{\boldsymbol I}$ is the reconstruction, and $\|\cdot\|_{\textsc{F}}$ denotes the Frobenius norm. The PSNR and SSIM measures shown in Figure~\ref{sdpr_image_recon}(d) increase with $\frac{m}{n}$ and indicate a good quality of reconstruction. An example reconstruction for $\frac{m}{n}=20$ is shown in Figure~\ref{sdpr_image_recon}(b) and the reconstruction error is shown in Figure~\ref{sdpr_image_recon}(c). The results show that the BPR algorithm is capable of retrieving the phase accurately.
\section{Conclusions}
We have addressed the problem of phase retrieval from oversampled binary measurements and demonstrated accurate reconstruction both in the presence and absence of noise. The optimization problem is formulated by amalgamating the principle of lifting with that of consistent  recovery, enforced by means of a one-sided quadratic loss function. One can also solve the BPR problem as generalized LASSO \cite{plan_yaniv, hassibi}, ignoring the nonlinearities in the measurement. However, the performance guarantees developed in \cite{plan_yaniv, hassibi} cease to apply as they require  the entries of $\boldsymbol A_i$ to be Gaussian, which does not hold in the case of BPR (since $\boldsymbol A_i=\boldsymbol a_i \boldsymbol a_i^\top$). The proposed BPR algorithm is iterative, based on APGD, and achieves an SRER of nearly $25$ dB for $20$ times oversampling. For images, the PSNR is as high as $30$ dB and the SSIM is about $0.75$. We have also considered the effect of noise and derived the CRB. The BPR algorithm is also robust to noise and lies within $2$ to $3$ dB of the CRB although it was not particularly optimized to combat noise. Relaxing the consistency criterion appropriately based on the noise level might lead to robustness -- this aspect requires further investigation.
\appendices{}
\section{Cram\'er-Rao Bound}
\label{crb_derivation_binary}
We derive the CRB for the binary measurements in \eqref{measurement_eq_binary_noisy}, corresponding to a fixed set of sensing signals $\left\{\boldsymbol a_i\right\}_{i=1}^{m}$. Related works in which CRBs were derived for PR are in the context of Gaussian noise corrupting the quadratic measurements \cite{balan1}, non-additive Gaussian noise prior to computing the quadratic measurement \cite{balan2}, uniformly distributed additive noise arising out of high-rate quantization \cite{pr_crb_fpp}, frame-based measurements \cite{bandeira}, and Fourier measurements \cite{cederquist}. In contrast to these works, our focus is on the extreme case of binary quantization, where none of the previously derived bounds hold.\\
\indent The measurement in \eqref{measurement_eq_binary_noisy} has the probability mass function
\begin{equation*}
p\left(y_i\right)=\left(1-\Phi\left(\tau-\left| \boldsymbol a_i^\top \boldsymbol x^*  \right|^2\right)\right)^{\bar{y}_i} \left(\Phi\left(\tau-\left| \boldsymbol a_i^\top \boldsymbol x^*  \right|^2\right)\right)^{1-\bar{y}_i},
\end{equation*}
where $\bar{y}_i = \frac{1+y_i}{2}$, $y_i \in \{-1, +1\}$, and $\Phi$ is the cumulative distribution function (c.d.f.) of noise. The log-likelihood function corresponding to the measurement vector $\boldsymbol y = [y_1, y_2, \cdots, y_m]$ is given by
\begin{eqnarray}
p_{\log}\left(\boldsymbol x^*\right)=\sum_{i=1}^{m}\bar{y}_i \log \left(1-\Phi\left(v_i\right)\right)+\left(1-\bar{y}_i\right)\log \left(\Phi\left(v_i\right)\right),
\label{ml_func_step1}
\end{eqnarray}
where $v_i=\tau-u_i^2$, with $u_i=\boldsymbol a_i^\top \boldsymbol x^*$. Differentiating both sides of \eqref{ml_func_step1} with respect to $\boldsymbol x^*$ gives
\begin{eqnarray}
\nabla p_{\log}\left(\boldsymbol x^*\right)=\sum_{i=1}^{m}\bar{y}_i \frac{2u_i \Phi'\left(v_i\right)}{1-\Phi\left(v_i\right)}\boldsymbol a_i-\left(1-\bar{y}_i\right) \frac{2u_i \Phi'\left(v_i\right)}{\Phi\left(v_i\right)}\boldsymbol a_i.
\label{ml_func_step2}
\end{eqnarray}
Indeed the regularity condition $\mathbb{E}_{\boldsymbol y}\left[\nabla p_{\log}\left(\boldsymbol x^*\right)\right]=\boldsymbol 0$, where $\mathbb{E}$ denotes the expectation, is satisfied, thereby guaranteeing existence of the CRB. Differentiating \eqref{ml_func_step2} again gives
\begin{eqnarray*}
\nabla^2 p_{\log}\left(\boldsymbol x^*\right)&=&\sum_{i=1}^{m}\bar{y}_i \frac{\left(1-\varphi_i\right) \left(2 \varphi_i'-4u_i^2 \varphi_i''  \right)-4u_i^2\varphi_i'^2}{\left(1-\varphi_i\right)^2}\boldsymbol A_i\nonumber\\&-& \left(1-\bar{y}_i\right) \frac{\varphi_i \left(2 \varphi_i'-4u_i^2 \varphi_i''  \right)+4u_i^2\varphi_i'^2}{\varphi_i^2}\boldsymbol A_i,
\label{ml_func_step3}
\end{eqnarray*}
where $\boldsymbol A_i = \boldsymbol a_i \boldsymbol a_i^\top, \varphi_i=\Phi\left(v_i\right)$, $\varphi'_i=\Phi'\left(v_i\right)$, and $\varphi''_i=\Phi''\left(v_i\right)$. The Fisher information matrix is given by  
\begin{equation}
\boldsymbol I_{\boldsymbol x^*} = -\mathbb{E}_{\boldsymbol{y}}\left[\nabla^2 p_{\log}\left(\boldsymbol x^*\right)\right]=\sum_{i=1}^{m}\frac{4u_i^2\varphi_i'^2}{\varphi_i\left(1-\varphi_i\right)} \boldsymbol A_i.
\label{beta_bar_eqn}
\end{equation}
The CRB for an unbiased estimate $\hat{\boldsymbol x}$ is given as $\text{Cov}\left(\hat{\boldsymbol x}\right)\succeq \boldsymbol I_{\boldsymbol x^*}^{-1}$. In the specific instance where the noise samples are i.i.d. Gaussian, as considered in Sections~\ref{bpr_measurement_noise_exp_sec} and \ref{np_crb_sec}, $\Phi$ and $\Phi'$ are the Gaussian c.d.f. and p.d.f., respectively.
\section{Descent Property of BPR With Projected Gradient Descent (PGD)}
\label{pgd_proof}
Consider the update rule of a projected gradient-descent (PGD) algorithm for BPR:
\begin{equation}
{\boldsymbol X}^{t+1}=\mathcal{P}_{\text{rank}-1}\left({\boldsymbol X}^{t}-\eta^t  \nabla \left. F\left( \boldsymbol X\right)\right|_{\boldsymbol X=\boldsymbol X^t}\right),
\label{update1}
\end{equation}
which can be rewritten as
\begin{equation}
{\boldsymbol X}^{t+1}=\arg \underset{\boldsymbol X \in \mathcal{R}_1}{\min}\text{\,\,}\frac{1}{2\eta^t}\left\|\boldsymbol X-\left({\boldsymbol X}^{t}-\eta^t  \nabla \left. F\left( \boldsymbol X\right)\right|_{\boldsymbol X=\boldsymbol X^t}\right)\right\|_{\textsc{F}}^2,
\label{update2}
\end{equation}
where $\|\cdot\|_\textsc{F}$ denotes the Frobenius norm and $\mathcal{R}_1$ is the set of all symmetric rank-1 matrices. Rearranging terms, the update turns out to be equivalent to
\begin{equation}
{\boldsymbol X}^{t+1}=\arg \underset{\boldsymbol X \in \mathcal{R}_1}{\min} P\left(\boldsymbol X,\boldsymbol X^t\right),
\label{update3}
\end{equation}
where $P\left(\boldsymbol X,\boldsymbol X^t\right)$ is defined as
\begin{eqnarray*}
P\left(\boldsymbol X,\boldsymbol X^t\right)&=& F\left( \boldsymbol X^t\right)+ \text{Tr}\left(\nabla  F\left( \boldsymbol X^t\right)^\top \left( \boldsymbol X -\boldsymbol X^t \right)\right)\\&+&\frac{1}{2\eta^t}\left\|\boldsymbol X -\boldsymbol X^t \right\|_{\textsc{F}}^2.
\end{eqnarray*} 
Suppose the gradient of $F\left( \boldsymbol X\right)$ is Lipschitz continuous (which we shall establish next), i.e., there exists a constant $L>0$ such that
\begin{equation*}
\left\| \nabla F\left( \boldsymbol X\right)-\nabla F\left( \boldsymbol Y\right) \right\|_{\textsc{F}} \leq L \left\| \boldsymbol X-\boldsymbol Y\right\|_{\textsc{F}},
\label{lipschitz}
\end{equation*}
for every pair of symmetric matrices $\left( \boldsymbol X,\boldsymbol Y\right)$. Then, for $\eta^t<\frac{1}{L}$, we have $F\left( \boldsymbol X\right)\leq P\left(\boldsymbol X,\boldsymbol X^t\right)$ for any symmetric $\boldsymbol X$, and, in particular, $F\left( \boldsymbol X^{t+1}\right)\leq P\left(\boldsymbol X^{t+1},\boldsymbol X^t\right)$. Since $\boldsymbol X^{t}$ and $\boldsymbol X^{t+1}$ belong to $\mathcal{R}_1$, we have that 
\begin{equation*}
F\left( \boldsymbol X^{t+1}\right)\leq P\left(\boldsymbol X^{t+1},\boldsymbol X^t\right) \stackrel{\text{(i)}}{\leq} P\left(\boldsymbol X^{t},\boldsymbol X^t\right)=F\left( \boldsymbol X^t\right),
\end{equation*}
where the inequality (i) is a consequence of \eqref{update3}. Therefore, the PGD algorithm reduces the objective provided that $F\left( \boldsymbol X\right)$ has a Lipschitz-continuous gradient. That $\nabla F\left( \boldsymbol X\right)$ is indeed Lipschitz continuous is established next.
\subsection{Lipschitz continuity of $\nabla F\left( \boldsymbol X\right)$} 
\begin{figure}[t]
\centering
$
\begin{array}{ccc}
\includegraphics[width=1.5in]{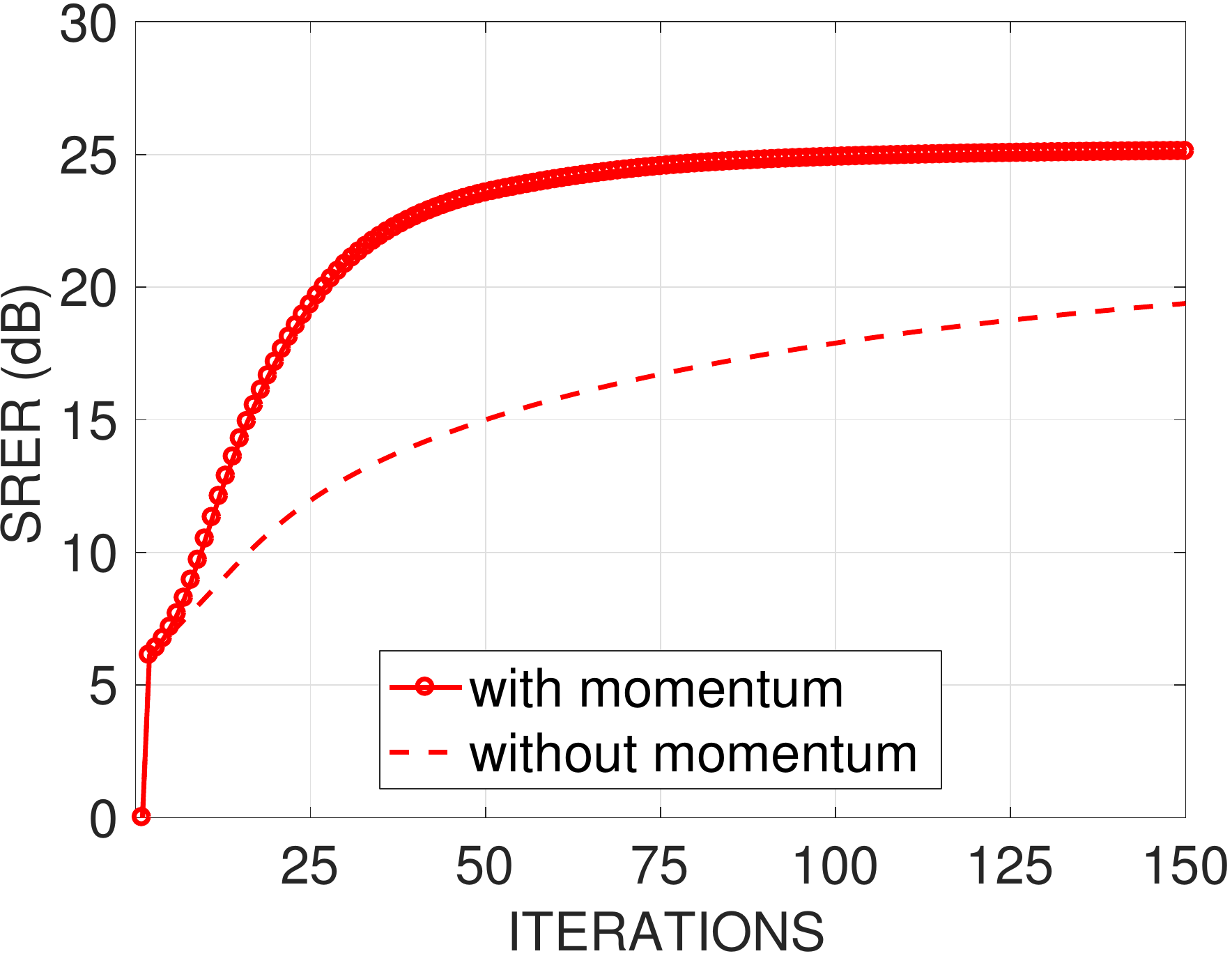}&
\includegraphics[width=1.5in]{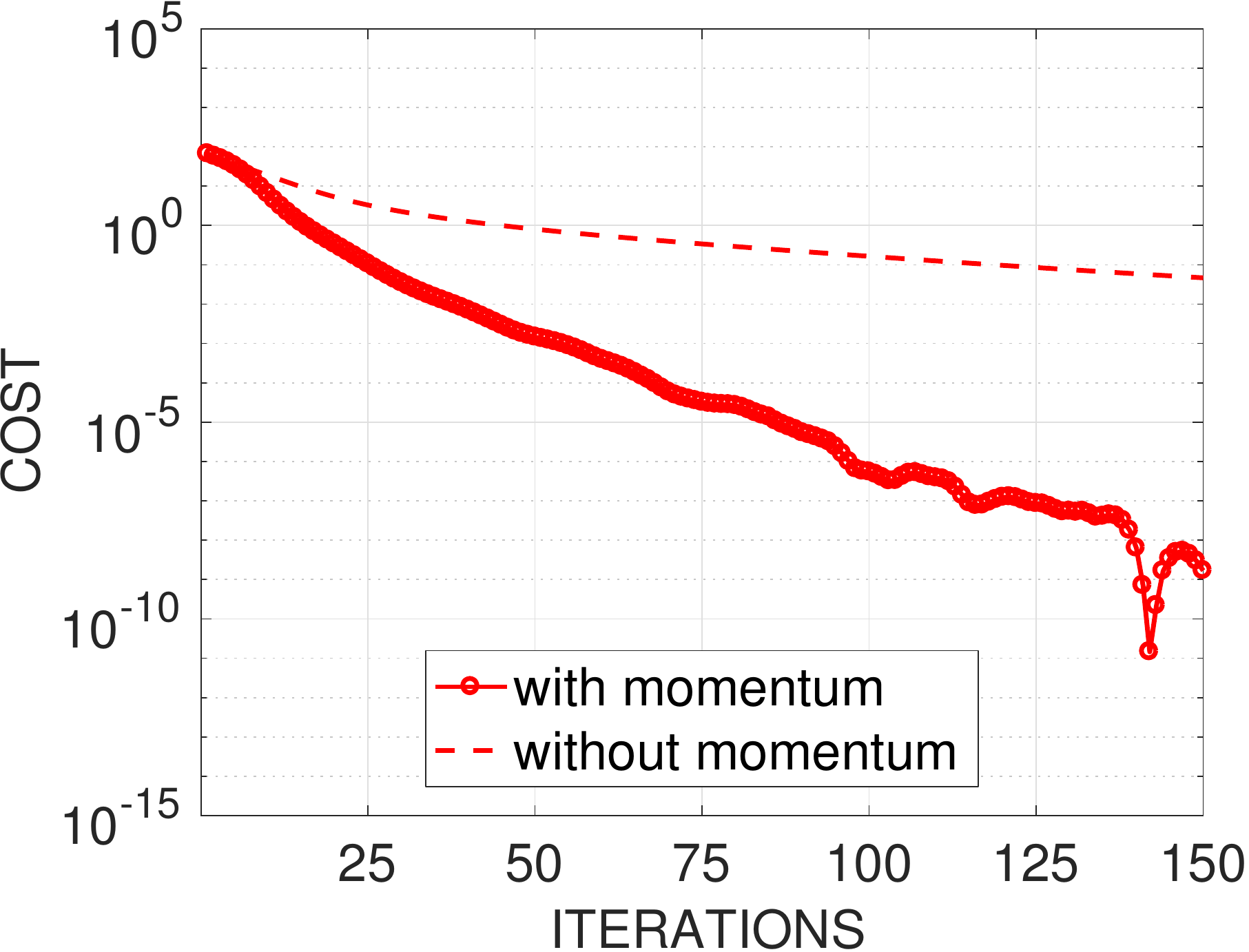}\\
\text{\small (a) SRER (dB) vs. iterations} & \text{\small (b) Cost } F(\boldsymbol X) \text{ vs. iterations}  \\
\end{array}
$
\caption{\small{(Color online) A comparison of the BPR algorithm implemented with and without the momentum factor.}}
\label{BPR_acceleration_effect_fig}
\end{figure}

Recall that 
\begin{equation*}
F\left(\boldsymbol X \right)=\sum_{i=1}^{m}f\left(y_i\left(\text{Tr}\left(\boldsymbol A_i \boldsymbol X\right) -\tau\right)\right).
\end{equation*}
For convenience, denote $u_i=y_i\left(\text{Tr}\left(\boldsymbol A_i \boldsymbol X\right) -\tau\right)$. The $(j_1,j_2)^{\text{th}}$ entry of the gradient $\boldsymbol G=\nabla F\left( \boldsymbol X\right)$ is given by
\begin{equation}
\boldsymbol G_{j_1,j_2}=\sum_{i=1}^{m}f'(u_i)y_i a_{i j_1}a_{i j_2},
\label{def_G}
\end{equation}
where $f'$ denotes the derivative of $f$. Differentiating \eqref{def_G} further with respect to ${\boldsymbol X}_{k_1,k_2}$, we get the Hessian (which is a tensor):
\begin{equation*}
\mathbb{H}_{j_1,j_2,k_1,k_2}=\sum_{i=1}^{m}f''(u_i) a_{i j_1}a_{i j_2} a_{i k_1}a_{i k_2},
\end{equation*}
after noting that $y_i^2=1$. The function $f''$ denotes the sub-differential of $f'$ and since $f(u)=\frac{1}{2}u^2\mathbbm{1}_{(u\leq0)}$, it follows that $f''(u_i)\leq 1$.\\
\indent For any positive-definite $\boldsymbol U\in \mathbb{R}^{n\times n}$, we have
\begin{eqnarray*}
&&\sum_{j_1,j_2=1}^{n}\sum_{k_1,k_2=1}^{n}{\boldsymbol U}_{j_1,j_2}\mathbb{H}_{j_1,j_2,k_1,k_2}{\boldsymbol U}_{k_1,k_2}\\ &=&\sum_{i=1}^{m}f''(u_i)\left(\boldsymbol a_i^\top \boldsymbol U \boldsymbol a_i \right)^2\\&\leq& \sum_{i=1}^{m}\left(\boldsymbol a_i^\top \boldsymbol U \boldsymbol a_i \right)^2\\&\leq& \lambda_{\max}^2\left(\boldsymbol U\right)\sum_{i=1}^{m}\left\| \boldsymbol a_i \right\|_2^4,
\end{eqnarray*}
where $\lambda_{\max}\left(\boldsymbol U\right)$ is the spectral norm or the largest eigenvalue of $\boldsymbol U$. Denoting $C_0=\displaystyle\sum_{i=1}^{m}\left\| \boldsymbol a_i \right\|_2^4$ and using the fact that the spectral norm is dominated by the Frobenius norm, we have $\lambda_{\max}^2\left(\boldsymbol U\right)\leq \left\| \boldsymbol U\right\|_{\textsc{F}}^2$, and therefore
\begin{eqnarray*}
\sum_{j_1,j_2=1}^{n}\sum_{k_1,k_2=1}^{n}{\boldsymbol U}_{j_1,j_2}\mathbb{H}_{j_1,j_2,k_1,k_2}{\boldsymbol U}_{k_1,k_2}  \leq C_0 \left\| \boldsymbol U\right\|_{\textsc{F}}^2,
\end{eqnarray*}
thereby establishing that $\nabla F\left( \boldsymbol X\right)$ is Lipschitz-continuous.\hfill $\blacksquare$\\
\indent Lipschitz continuity guarantees that the PGD algorithm for BPR does not increase the cost function in every iteration. This property is not guaranteed to hold when a momentum factor is added in every iteration, due to non-convexity of the rank-1 constraint. However, we have observed empirically that the incorporation of a momentum term does reduce the cost at a rate faster than the PGD scheme (cf. Figure~\ref{BPR_acceleration_effect_fig}). A similar observation was made my Cand\`es et al. in the context of PhaseLift (cf. Section 4.1 of reference [27]).    
\section{Binary Phase Retrieval With Fourier Measurements}
\label{bpr_dft_app}
\begin{figure}[t]
\centering
$
\begin{array}{ccc}
\includegraphics[width=2.in]{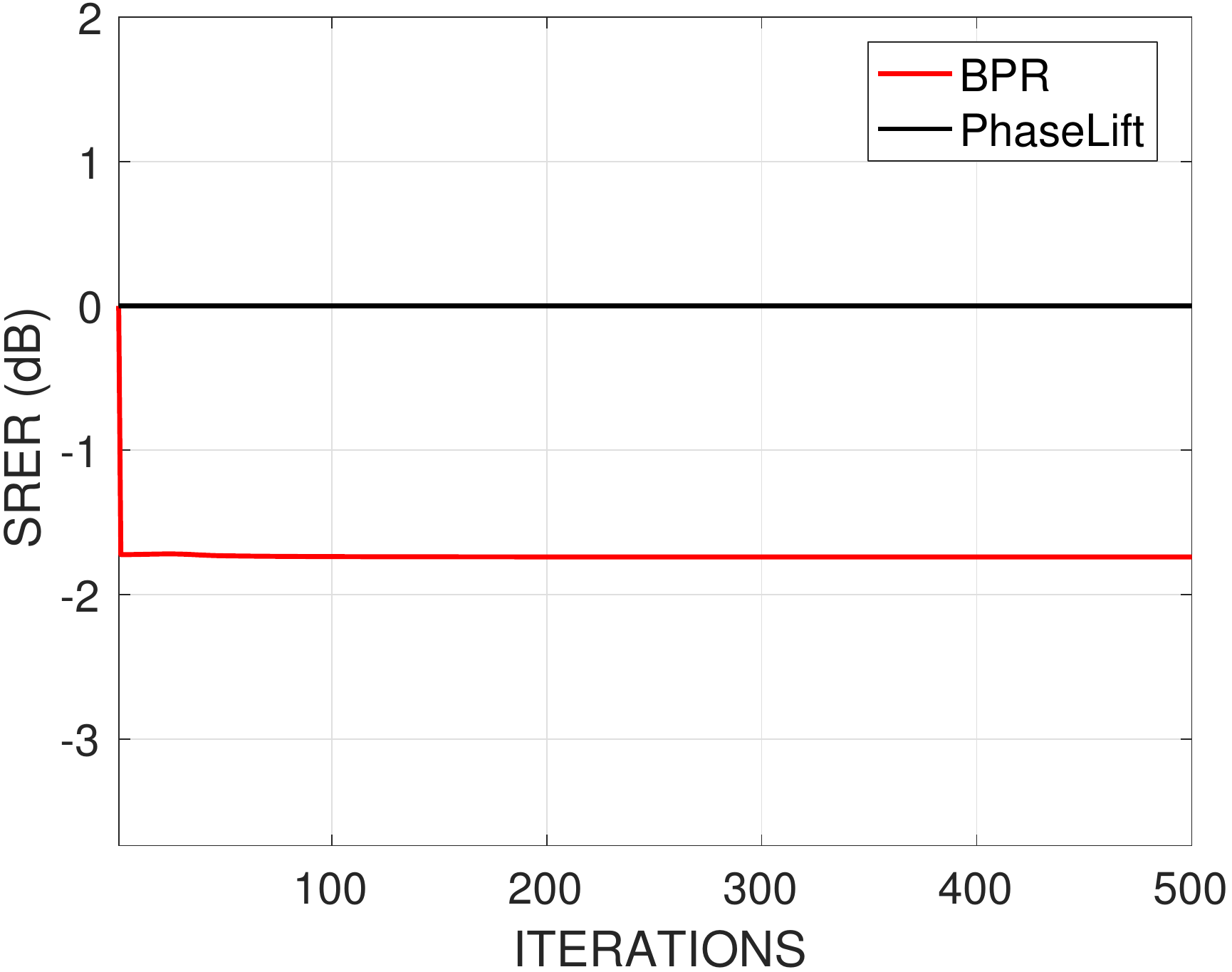}&
\end{array}
$
\caption{\small{(Color online) Recovery failure of BPR and PhaseLift corresponding to oversampled DFT magnitude measurements with no randomization (that is, ${\boldsymbol W}_j$s are not used). The oversampling factor is taken to be $\frac{m}{n}=20$.}}
\label{BPR_ovdft}
\end{figure}

\indent Here, we illustrate that the BPR algorithm is not restricted to Gaussian measurements and can be applied to Fourier measurements as well. We consider a Fourier sampling scheme of the structured illumination type, which was considered in the context of PhaseLift. In this setup, one considers the measurement matrix 
\begin{equation}
\boldsymbol A=\left[\begin{array}{cccc} \boldsymbol F  \boldsymbol W_1\\  \boldsymbol F  \boldsymbol W_2\\ \vdots\\  \boldsymbol F  \boldsymbol W_k\end{array}\right],
\label{Amatrix}
\end{equation}
where $\boldsymbol F$ is the $n \times n$ discrete Fourier transform (DFT) matrix, $\boldsymbol W_j$s are $n \times n$  diagonal matrices containing random binary entries (0 or 1 with probability $\frac{1}{2}$) on the diagonal, and $k=\frac{m}{n}$ is the oversampling factor. The measurements $|\boldsymbol{a}_i^\textsc{H}\boldsymbol x|^2$, where $\boldsymbol{a}_i$ is complex-valued and denotes the $i^{\text{th}}$ row of the $\boldsymbol{A}$ constructed as described in \eqref{Amatrix} above, are quantized as $\pm 1$, depending on whether they exceed a threshold $\tau$ or not. In this case, $|\boldsymbol{a}_i^\textsc{H}\boldsymbol x|^2 = \text{Tr}({\boldsymbol A_i}{\boldsymbol X})$, where ${\boldsymbol A_i} = \boldsymbol{a}_{i_{\text{re}}}\boldsymbol{a}_{i_{\text{re}}}^\top + \boldsymbol{a}_{i_{\text{im}}}\boldsymbol{a}_{i_{\text{im}}}^\top$, and $\boldsymbol{X}$ is real. The subscripts `re' and `im' denote the real and imaginary parts, respectively. The threshold $\tau$ is set according to the criterion described in Section III B. To recall, the threshold $\tau$ is set such that $\text{Prob}\left(|\boldsymbol{a}_i^\textsc{H}\boldsymbol x|^2 > \tau \right) = \text{Prob}\left(|\boldsymbol{a}_i^\textsc{H}\boldsymbol x|^2 < \tau \right) = \frac{1}{2}$. The reconstruction performance of BPR and PhaseLift for this setting is shown in Figure~\ref{BPR_fourier_fig1}. We observe that PhaseLift converges faster than BPR, but the SRER of BPR is about $4$ dB higher than that of PhaseLift after convergence.\\
\indent Instead of using the structured illumination model considered above, if one were to employ only oversampled DFT measurement matrices (without the randomizing $\boldsymbol W_j$s), both BPR and PhaseLift would fail to reconstruct the signal as illustrated in Figure~\ref{BPR_ovdft}.
\section{Settings for the Other Algorithms}
\label{other_algo_settings}
The settings for the competing algorithms are explained in the following. We are unable to include this discussion in the main manuscript due to the four-page constraint. 
\begin{itemize}
\item The PhaseLift algorithm minimizes the quadratic loss  
\begin{equation*}
Q\left(\boldsymbol X\right)=\sum_{i=1}^{m}\left(\text{Tr}\left( \boldsymbol A_i \boldsymbol X\right)-y_i\right)^2, 
\end{equation*}
where $\left\{y_i\right\}_{i=1}^{m}$ are the measurements, using an accelerated projected gradient algorithm. In the implementation, we choose the step-size parameter following the exact line-search procedure, for which a closed-form expression can be calculated as follows:
\begin{equation}
\eta^t_{\text{PhaseLift}}=\frac{\sum_{i}^{m}\left(\text{Tr}\left(\boldsymbol A_i \boldsymbol X^t\right)-y_i \right)\text{Tr}\left( \boldsymbol A_i \boldsymbol G^t \right)}{\sum_{i}^{m}\left( \text{Tr}\left( \boldsymbol A_i \boldsymbol G^t \right)\right)^2},
\label{pl_ls_eq}
\end{equation}
where $\boldsymbol G^t=\nabla Q\left(\boldsymbol X\right)\big|_{\boldsymbol X=\boldsymbol X^t}$.
\item For TWF, we employed the implementation available on the authors' website\footnote{\url{http://web.stanford.edu/~yxchen/TWF/}.}. The TWF routine accepts the measurements $y_i$ and the sampling vectors $\boldsymbol a_i$ for reconstruction, and returns an estimate $\hat{\boldsymbol x}$ of the ground-truth. The implementation assumes a Poisson likelihood on the measurements, leading to the loss function 
\begin{eqnarray*}
L(\boldsymbol x)&=&\sum_{i=1}^{m}\ell\left(y_i,\left| \boldsymbol a_i^\top \boldsymbol x \right|^2\right)\\&=&\sum_{i=1}^{m}y_i \log\left(\left| \boldsymbol a_i^\top \boldsymbol x \right|^2\right)-\left| \boldsymbol a_i^\top \boldsymbol x \right|^2.
\end{eqnarray*}
\item AltMinPR is implemented exactly following the AltMinPhase algorithm proposed in \cite{netrapalli}.   
\end{itemize}
\section{A Comparison of Run-times}
\label{run_time_sec}
The per-iteration run-times of the algorithms under consideration are given in Table \ref{table_run_time}. All algorithms are implemented on  MATLAB-2016b platform, running on a Mac-OS 11.06 computer having 8 GB RAM and 3.2 GHz Intel Core i5 processor. The asymptotic per-iteration complexity of BPR is the same as that of PhaseLift and AltMinPR, namely $\mathcal{O}\left(n^3\right)$, where $n$ is the dimension of the ground-truth vector. BPR has the overhead of step-size selection, which is done numerically using a grid search and accounts for a large portion (nearly $99.7\%$) of the per-iteration run-time as shown in Table \ref{table_run_time}. In contrast, the optimal step-size for exact line-search in case of PhaseLift can be computed in closed-form as shown in \eqref{pl_ls_eq} --- this takes an order of magnitude lesser time than a grid search.

\begin{center}
\begin{table*}[!h]
\centering
\begin{tabular}{c |c| c}
\hline\hline
Algorithm & Total per-iteration run-time & Per-iteration line-search time \\
&  (in seconds) & (in seconds)\\
\hline
BPR & $0.6318$ & $0.6304$  \\

\hline
PhaseLift &$0.0689$ & $0.0682$ (Closed-form) \\

\hline
TWF &$1.93\times 10^{-4}$ & Not applicable  \\

\hline

AltMinPR &$9.90\times 10^{-3}$ & Not applicable  \\
\hline\hline

\end{tabular}
\vspace{0.1in}
\caption{\small{A comparison of run-time per iteration corresponding to various algorithms. The total per-iteration run-time is inclusive of line-search, in case of BPR and PhaseLift.}}
\label{table_run_time} 
\end{table*}
\end{center}

\ifCLASSOPTIONcaptionsoff
  \newpage
\fi

\end{document}